\documentclass{elsart}
\usepackage{graphicx,amssymb}
\journal{Physics Letters A}
\begin{document}
\begin{frontmatter}

\title{Complex Optical Potentials and Pseudo-Hermitian Hamiltonians}
\author[Darj]{Ram Narayan Deb},
\author[Inst]{Avinash Khare},
\author[Snb]{Binayak Dutta Roy\corauthref{cor}}
\corauth[cor]{Corresponding author.}
\ead{bnyk@bose.res.in}

\address[Darj]{Department of Physics, Darjeeling Government College,
Darjeeling, West Bengal, India.}

\address[Inst]{Institute of Physics, Sachivalay Marg, Bhubaneswar, Orissa, India.}

\address[Snb] {Satyendra Nath Bose National Centre for 
Basic Sciences,
Block-JD, Sector-III, Salt Lake,
Calcutta - 700098, India.}

\begin{abstract}
Recently some authors have broadened the scope of canonical
quantum mechanics by replacing the conventional Hermiticity
condition on the Hamiltonian by a weaker requirement 
through the introduction of the notion of pseudo-Hermiticity.
In the present study we investigate eigenvalues, transmission
and reflection from complex optical potentials enjoying the property of
pseudo-Hermiticity.
\end{abstract}
\end{frontmatter}

 Bender, his collaborators and others [1-23]
 in a series of papers,
investigated some non-Hermitian Hamiltonians which violate parity $(P)$ and time
reversal $(T)$ symmetry but are $PT$ invariant. These systems comprised of a
particle moving in a complex potential, the real part of which is parity-even
while the imaginary part is odd. Through various examples it was found that the
energy eigenvalues were real and bounded from below. Though in some cases, for a
range of parameters (contained in the potential), complex eigenvalues do occur,
but these are associated with a spontaneous breaking of $PT$ symmetry.
Subsequently Mostafazadeh \cite{24} has provided a basic mathematical setting
for such systems by introducing the notion of pseudo-Hermiticity of the 
Hamiltonian $H$ via the condition $H^{\dagger}={\eta}H{\eta}^{-1}$ where
${\eta}$
is a Hermitian, linear and invertible operator. It is suitable to define an
indefinite inner product of two state vectors: ${\ll}{\psi}_{1}{\vert}{\psi}_
i{2}{\gg}_{\eta}$ = ${\langle}{\psi}_{1}{\vert}{\eta}{\vert}{\psi}_{2}{\rangle}$
which is time translationally invariant with $\eta$ playing the role of a 
metric operator in Hilbert space.
 One can then go on to show that the eigenvalues of $H$ are
either real or occur as complex conjugate pairs and the eigenvectors constitute
a complete biorthonormal system. For the examples considered by Bender and 
others $\eta$ is the parity operator $(\eta=P)$. 

   In the present study we attempt to put such Hamiltonians in a more 
`physical' setting and moreover extend the discussion to continuum states as
well. We encounter some rather amusing features. We recall \cite{25} the 
optical (or cloudy crystal ball) model of the atomic nucleus, where the 
interaction of a neutron (or some other projectile) with the nucleus is 
described through a complex potential $V+iW$, where $V$ and $W$ are its real
and imaginary parts respectively. From the continuity equation obeyed by the 
probability and probability current densities (following from Schr\"odinger
equation) it is easily seen that non-vanishing $W$ implies a sink (or source)
for the probability, depending on the sign of $W$, which is taken to 
correspond to absorption (or emission) of particles with respect to the 
incident beam, and this furnishes a rather useful phenomenological description
of elastic scattering in the presence of open inelastic channels. In one of the
versions of this picture, known as the surface absorption model, the imaginary
part $(W)$ of the potential is taken to be proportional to the spatial 
derivative (gradient) of the real part $(V)$. We shall see that a 
one-dimensional `cartoon' of the surface absorption version of the optical 
model
leads naturally to a $PT$-invariant (or more appropriately a pseudo-Hermitian)
Hamiltonian. We illustrate some novel features of pseudo-Hermitian Hamiltonians
through two examples so chosen such that the results are expressible in terms 
of elementary functions. 

           Consider a particle of mass $m$ moving in one dimension under the
influence of a potential the real part $(V)$ of which is an attractive square
well of depth $V_{0}$ and range $a$ and whose imaginary part $(W)$ is 
proportional to the derivative of the real part. Accordingly the motion is 
governed by the Hamiltonian
\begin{equation}
H= \frac{p^{2}}{2m}+V+iW=\frac{p^{2}}{2m}-V_{0}\theta(\frac{a}{2}-{\vert}
x{\vert})+i{\lambda}[\delta(x-\frac{a}{2})-\delta(x+\frac{a}{2})].
\end{equation}  
  Note that the Hamiltonian is not parity invariant as $V+iW{\longrightarrow}
V-iW$ ($W$ being an odd function of $x$ ), nor is it time reversal invariant as 
the operation of complex conjugation takes $V+iW{\longrightarrow}V-iW$. 
However the system enjoys symmetry under the combined $PT$ transformation. More
relevantly the Hamiltonian is pseudo-Hermitian viz. $H^{\dagger}=PHP^{-1}$ with
the parity operator $P$ being allowed to play the role of
 the `metric' operator $(\eta)$ in the Hilbert space.

 Analysing the bound state situation to begin with, it is convenient to put the
energy eigenvalue $E=-B$ with $B>0$ and to define $\frac{2mB}{{\hbar}^{2}}=
{\beta}^{2}$, so that ${\psi}_{I}(x)=A{e^{{\beta}x}}$ for $x<-{\frac{a}{2}}$
(as $\psi$ must vanish as $x{\longrightarrow}-{\infty}$) and
$\psi_{III}(x)=D{e^{-{\beta}x}}$ in the region $x>+\frac{a}{2}$ (as $\psi$ must
vanish as $x{\longrightarrow}+{\infty}$). In the range ${\vert}x{\vert}<
\frac{a}{2}$ we have the solution $\psi_{II}(x)=C_{1}e^{iqx}+C_{2}e^{-iqx}$
where $q^{2}=\frac{2m(V_{0}-B)}{\hbar^{2}}$. Implementing the continuity and
jump conditions $ \psi_I(\frac{-a}{2}) = \psi_{II}(\frac{-a}{2})$,
$\psi_{II}(\frac{+a}{2}) = \psi_{III}(\frac{+a}{2})$,
$\psi_{II}^{\prime}({\frac{-a}{2}})-\psi_{I}^{\prime}({\frac{-a}{2}}) = 
 -i\tilde{\lambda}{\psi_I}({\frac{-a}{2}})$,
$ \psi_{III}^{\prime}(\frac{+a}{2})-{\psi_{II}}^{\prime}({\frac{+a}{2}}) = 
i\tilde{\lambda}{\psi_{III}}({\frac{+a}{2}})$
 where $\tilde{\lambda}~~~{\equiv}~~~{\frac{2m{\lambda}}{\hbar^{2}}}$, we 
arrive at the eigenvalue condition

\begin{equation}
(q^{2}-{\beta^{2}}-{\tilde{\lambda}^{2}})~~Sin(qa)=2{\beta}q~~Cos(qa).
\end{equation}

Before going on to a discussion of the solution to this equation, it is 
revealing to ask the question: Is there a minimum depth $V_{0}$ below which
a bound state (with the usual meaning of the concept as elaborated below)
does not exist ? 
For the square well potential (with $\lambda = 0$) we know that there is at 
least one bound state howsoever weak the attraction ( given by $V_0$ )
may be. Also for $\sqrt{\frac{2mV_0a^{2}}{\hbar^{2}}} < \frac{\pi}{2}$ there is
only one bound state. Let us choose the real part of the potential to lie in
this region ( say $\frac{2mV_0a^2}{\hbar^2} = 1$ ) and look for the effect of 
the imaginary part ($\lambda$) on the binding energy by numerically solving
eq.(2). The result is shown graphically in $Fig. 1(a)$. Observe that the 
binding energy goes to zero at 
$\tilde\lambda \equiv \frac{2m\lambda}{\hbar^2} = 1$. This is readily seen by
looking at the zero binding condition, namely, $\beta = 0$ in the eigenvalue
equation [eq. (2)] which yields the relation
\begin{equation}
(V_{0}-{\frac{2m{\lambda^2}}{\hbar^{2}}})
Sin({\sqrt{\frac{2mV_{0}}{\hbar^{2}}}}a)=0.
\end{equation}
The relevant result for this minimum depth is thus
\begin{equation}
 V_{0}=\frac{2m{\lambda}^{2}}{\hbar^{2}}.
\end{equation}
In this limiting situation the wavefunction corresponding to zero binding
$(B=0)$ is

${\psi_{I}}=A$ ${\;\;\;}$ for $x<-{\frac{a}{2}}$,

${\psi_{II}}=Ae^{-i{\frac{m{\lambda}a}{\hbar^{2}}}-i\frac{2m{\lambda}x}{\hbar
^{2}}}$ ${\;\;\;}$
for -${\frac{a}{2}}< x <+ {\frac{a}{2}}$

${\psi_{III}}=Ae^{-i{\frac{2m{\lambda}a}{\hbar^{2}}}}$ ${\;\;\;}$
 for $x>+{\frac{a}{2}}$.

To examine analytically what happens when the strength of imaginary part 
crosses the critical value, let us choose for simplicity
$\sqrt{\frac{2mV_0a^2}{\hbar^2}} = \frac{\pi}{2}$ and look in the neighbourhood
of the point ${\tilde\lambda}^{2} = \frac{2mV_0}{\hbar^2} = {\omega}^2$ viz
${\tilde\lambda}^2a^2 = {\omega}^2a^2 - \epsilon$ where, with $\epsilon$ 
positive, a bound state should exist. With sufficiently small $\epsilon$ the 
roots of the transcendental equation [eq. (2)] reduces to finding the zeros of
a cubic form and leads us to three solutions 
${\beta}a = -\frac{3}{8}\epsilon \pm \sqrt{\frac{\epsilon}{2}} + O({\epsilon}^2)
$ and ${\beta}a = -2 + O(\epsilon)$. In order to have 
$\psi(x)= e^{-{\beta}x} \longrightarrow 0 $ as $ x \longrightarrow \infty $  and
$\psi(x)=e^{{\beta}x} \longrightarrow 0$ as $x \longrightarrow -{\infty}$ 
corresponding to bound states it is necessary to have $\beta > 0 $ and thus 
only one of these three roots is physical namely 
${\beta}a = -\frac{3}{8}\epsilon + \sqrt{\frac{\epsilon}{2}} + O({\epsilon}^2)$
, the other two being unphysical. With $\tilde\lambda$ values beyond the 
critical value looking again at the neighbourhood of that point we obtain
${\beta}a = \frac{3}{8}|\epsilon| \pm i \sqrt{\frac{|\epsilon|}{2}}$ and the 
third root at $-2 + O(\epsilon)$ which as before is inadmissible. While the 
wavefunctions corresponding to the complex conjugate roots are well behaved in
 the sense that $|\psi(x)|^{2} \longrightarrow 0$ as $x \longrightarrow 
\pm \infty$ and are square integrable, nevertheless there is a serious 
difficulty regarding the physical interpretation of the time dependence of the
 probablity. Thus while the wavefunction corresponding to the root with 
negative imaginary part can be thought of as a decaying state, its inevitable
partner with positive imaginary part grows with time and is thus physically
unacceptable. Moreover these two states are not orthogonal in the sense that
$\ll \psi_{1}|\psi_{2} \gg_{\eta} = \langle\psi_{1}|{\eta}|\psi_{2}\rangle
\ne 0 $ (where $\eta = P$ ) while 
$\ll \psi_{1}|\psi_{1} \gg_{\eta} = 0 = \ll \psi_{2}|\psi_{2} \gg_{\eta}$. Thus
the system itself ceases to have usual physical significance when these 
complex conjugate roots make their appearence, namely, when the strength of the
imaginary part of the potential $\lambda$ exceeds the critical value.
It may be noted more generally that if the real part of the potential were 
deeper so as to support more than one bound state then the boundstates continue to be real (and interpretable) as long as the imaginary part is less than
the critical value at which the least bound state complexifies and the system
no longer sustains its physical meaning. It may also be noted that in the 
region of parameter space when complex energy eigenvalues occur, the 
wavefunctions are given by 
$\psi(x) \sim e^{\frac{3}{8}|\epsilon|\frac{x}{a}+
i\sqrt{\frac{|\epsilon|}{2}}\frac{x}{a}}$ for $x < 0$  and
$\psi(x) \sim e^{-{\frac{3}{8}}|\epsilon|\frac{x}{a}
-i\sqrt{\frac{|\epsilon|}{2}}\frac{x}{a}}$ for $x > 0$ 
(for ${\tilde\lambda}^{2}$
slightly greater than the critical value $\frac{2mV_{0}}{\hbar^2}$)
[as also the other root giving $\psi(x) \sim e^{\frac{3}{8}|\epsilon|
\frac{x}{a}-i\sqrt{\frac{|\epsilon|}{2}}\frac{x}{a}}$ for $x < 0$ and
$\psi(x) \sim e^{-\frac{3}{8}|\epsilon|\frac{x}{a}+
i\sqrt{\frac{|\epsilon|}{2}}\frac{x}{a}}$ for $x > 0$]. These
are not $PT$-invariant
and as such with the Hamiltonian enjoying the symmetry and the states violating
it, we see that $PT$ is spontaneously broken.

 This situation may be contrasted with the system of a particle moving in a 
complex Morse potential in one dimension which was so contrived as to have only
real energy eigenvalues [26]. It has also been pointed out [27] that for this 
potential the resulting Hamiltonian is $\eta$-pseudo-Hermitian 
with $\eta = e^{-{\theta}p}$ where $p$ is the momentum operator. As has been
shown by Mostafazadeh [28] the necessary and sufficient condition that a 
pseudo-Hermitian Hamiltonian has only real eigenvalues is that we may write
$\eta = A^{\dagger}A$, where $A$ is a linear and invertible operator. Clearly
for the operator $e^{-{\theta}p}$ the corresponding $A$ is 
$e^{-\frac{1}{2}{\theta}p}$. Noting that this is not the case with the examples
chosen here, we have real eigenvalues only for a certain regime of parameters.
               
 A convenient framework for the discussion of transmission and reflection for
the problem at hand is provided by the $S-$matrix approach. We introduce 
asymptotic channel states  
 $\vert m, k \rangle$, with $m= R, L$ : 
$\langle x \vert R, k \rangle = e^{ikx}$ and 
$\langle x \vert L, k \rangle = e^{-ikx}$,
where $R$ and $L$ stand for right moving and left moving free particle states
and the wave-number  $k = \sqrt {{2m\over \hbar^2}E}$. One can then define the
$S$ operator whose on-shell matrix elements  
$\langle m,k\vert S \vert n,k \rangle =
S_{m,n}(k)$ gives the probability amplitude for a state starting off in the 
remote past as ${\vert}n,k{\rangle}$, to be found, as a result of evolution
through the interaction with the potential, in the state ${\vert}m,k{\rangle}$
in the remote future. This $2\times2$ matrix $S$ would have been unitary if the 
potential were real, but in the present case this will not be so. With the use
of these conventions for enumerating channels, it is evident that the $S$
matrix elements are related to the familiar transmission ($t_{R}$ and $t_{L}$)
and reflection ($r_{R}$ and $r_{L}$) amplitudes for right and left travelling
particles (indicated through the subscripts): 
\begin{eqnarray}                          
{ {S_{RR}~~~~  S_{RL}}\choose {S_{LR}~~~~ S_{LL}} } 
={ {t_R~~~~ r_L}\choose
    { r_R~~~~   t_L} }
\end{eqnarray}
With hermitian Hamiltonians the states evolve in a unitary manner and the 
$S-$matrix obeys the unitarity condition $S^{\dagger}S=I$ which in this case 
would imply the relation $|t_{R}|^{2} + |r_{R}|^{2} = 1$, 
$|t_{L}|^{2} + |r_{L}|^{2} = 1$ and $ t_{R}^{\ast}r_{L} + r_{R}^{\ast}t_{L}
 = 0$. The first two of these conditions are nothing but the conservation of
 probability for the right and left incident beams respectively while the third
describes the phase relationships. Note that in a left-right symmetric 
situation this reduces to $t^{\ast}r + r^{\ast}t = 0$ or that the transmission
and reflection amplitudes are out of phase by $\frac{\pi}{2}$.
In our case since the Hamiltonian is P-pseudo hermitian, it is clear that
as $H^{\dagger} = PHP^{-1}$
the $S-$matrix obeys the pseudo unitarity condition 
$P^{-1}S^{\dagger}PS = I$. 
The $P$ operation in the $|L\rangle$ , $|R\rangle$ basis in which we have 
expressed the $S-$ matrix is given by the matrix
\begin{eqnarray}
P &=& {{0~~~~~~1}\choose{1~~~~~~0}}
\end{eqnarray}
This implies that
$t_{L}^{\ast}t_{R} + r_{L}^{\ast}r_{R} = 1$, $r_{L}^{\ast}t_{L} + t_{L}^{\ast}
r_{L} = 0$ and $r_{R}^{\ast}t_{R} + r_{R}t_{R}^{\ast} = 0$. The last two 
conditions imply that the reflection and transmission amplitudes are out of 
phase by $\frac{\pi}{2}$ for both the left and right incident beams.

For the particular model defined by the Hamiltonian given in eq.(1),   
the transmission amplitudes for left and right incident beams are explicitly 
found to be
the same, and in our example is given by
\begin{equation}
t_R = t_L = {2 q k e^{-ika} \over 2 q k{\;} Cos (qa) -i{\;} (q^2 + k^2 - 
{\tilde\lambda}^2){\;} Sin (qa)} .
\end{equation}  
The fact that $t_{R}=t_{L}$ follows from $PT$ symmetry of the potential. 
Since the Hamiltonian is $PT-$ symmetric, hence the $S-$matrix is also 
$PT-$ symmetric. For a hermitian Hamiltonian which is invariant under the
time-reversal it is well known (see for example the discussion in the text-book
by A.S. Davydov [29]) that the $S-$ matrix gets transposed under the operation
of time reversal $(T)$. Following the same procedure it can be seen that even 
though, as in our case, the Hamiltonian changes under 
time reversal $(T)$ and parity $(P)$
separately , but is
symmetric under their joint operation (that is $PT$)
 the $S-$ matrix suffers a transposition under $T$. Accordingly 
$(PT)H(PT)^{-1} = H$ while $PHP^{-1} = H^{\dagger}$ ; and
thus $THT^{-1} = PHP^{-1} = H^{\dagger}$. Introducing a unitary 
operator $O$ ($O^{\dagger}O=I$) 
such that $OH^{\ast} = H^{\dagger}O$ then following Davydov [29] we 
can assert that $T = OK$, where $K$ is the operator for complex conjugation.
 Thus the 
wavefunction corresponding to the time reversed state is
$\psi_{-a} = T\psi_{a} = OK\psi_{a} = O\psi_{a}^{\ast}$.
The $S-$  matrix element between the time reversed states is

$~~~~~~~~~~~~~S_{-a,-b} = \langle\psi_{-a}|S|{\psi_{-b}} \rangle
= \langle T\psi_{a}|S| T{\psi}_{b}{\rangle} 
=\langle O\psi_{a}^{\ast}|S|O{\psi_{b}}^{\ast}
\rangle\\ 
~~~~~~~~~~~~~~~~~~~~~~~~~~~=\langle \psi_{a}^{\ast}|O^{\dagger}SO|
\psi_{b}^{\ast}\rangle $
$=\langle \psi_{a}^{\ast}|(S^{\dagger})^{\ast}|
\psi_{b}^{\ast}\rangle
=S_{ba}$\\
where we have used the relation $OH^{\ast} = H^{\dagger}O$ to obtain
$O^{\dagger}SO = (S^{\dagger})^{\ast}$ and then the fact that
$\langle \psi_{a}^{\ast}|(S^{\dagger})^{\ast}|\psi_{b}^{\ast}\rangle = 
\langle \psi_{a}|S^{\dagger}|\psi_{b}\rangle^{\ast} =
\langle \psi_{b}|S|\psi_{a}\rangle = S_{ba}$.

Therefore the $S-$matrix get transposed under the operation of time reversal(T)
. Now applying the parity transformation eq.(6) 
to the time reversed $S-$matrix we get
\begin{eqnarray}
(PT)S(PT)^{-1} &=& {{S_{LL}~~~~~~S_{RL}}\choose {S_{LR}~~~~~~S_{RR}}}
\nonumber
\end{eqnarray}
``Demanding the invariance of the $S-$matrix under $PT-$ transformation" we get
$S_{LL} = S_{RR}$ implying $t_{L} = t_{R}$. It may also be remarked that in the
case when the Hamiltonian is both parity and time reversal invariant 
(separately) then $P$ invariance alone implies $t_{R} = t_{L}$ and
$r_{R} = r_{L}$, while $T$ invariance alone leads to $t_{R} = t_{L}$. This may
be compared to what is obtained in the case where we have 
violation of $P$ and $T$ with $PT$ conservation.
 
The reflection amplitudes, however, for left and right moving particles are 
different and are given by 
\begin{equation}
r_L = i{\;}{[q^2 - (k + {\tilde\lambda})^2] Sin (qa){\;} e^{-ika}
\over 2 q k{\;} Cos (qa) - i{\;}(q^2 + k^2 - {\tilde\lambda}^2){\;}Sin (qa)} . 
\end{equation}

\begin{equation}
r_R = i{\;}{[q^2 - (k - {\tilde\lambda})^2] Sin (qa){\;} e^{-ika}
\over 2 k q{\;} Cos (qa) - i{\;} (q^2 + k^2 - {\tilde\lambda}^2){\;}Sin (qa)} .
\end{equation}

\begin{figure}[t]
\begin{center}
\includegraphics*[width=2in,angle=270]{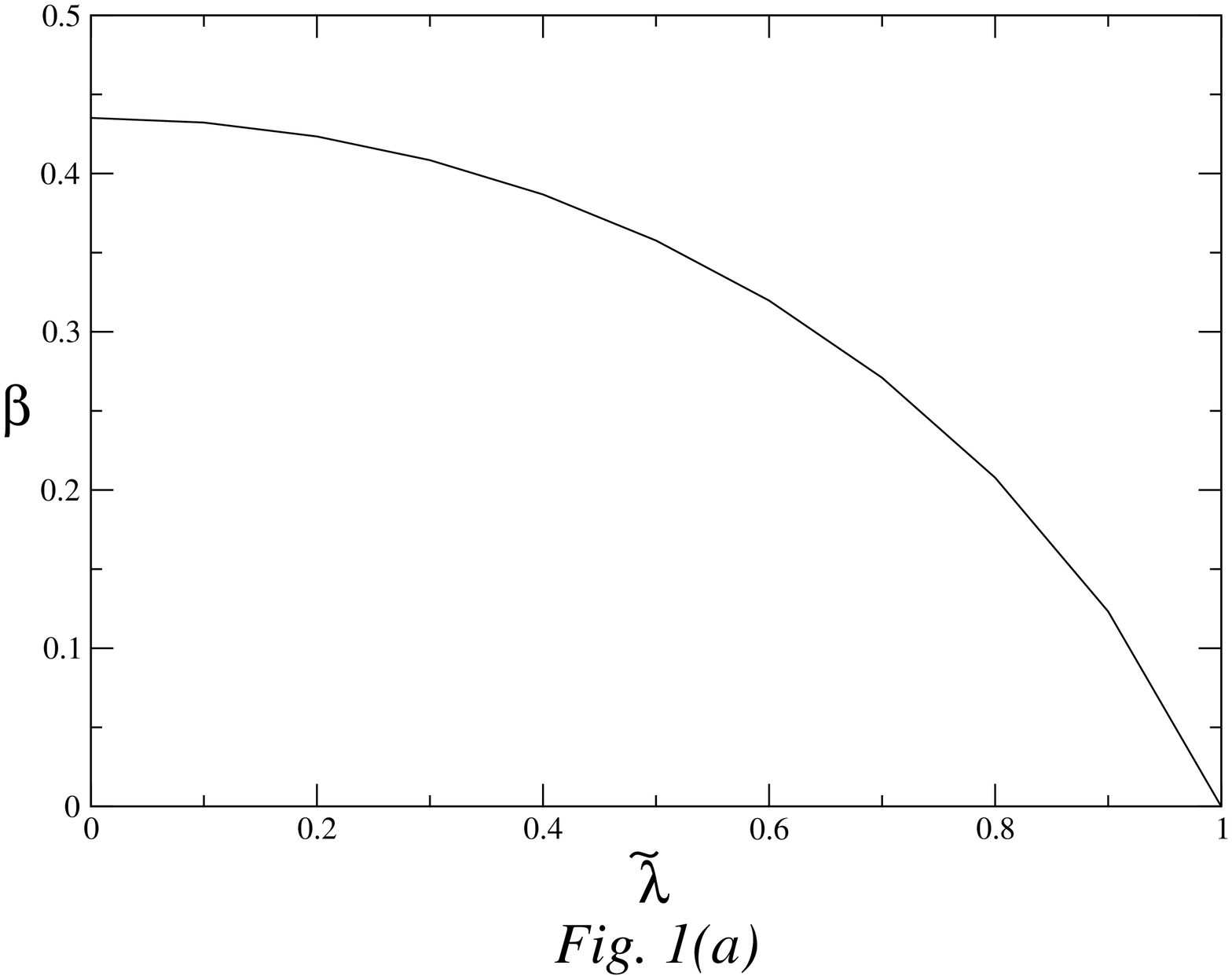}
\includegraphics*[width=2in,angle=270]{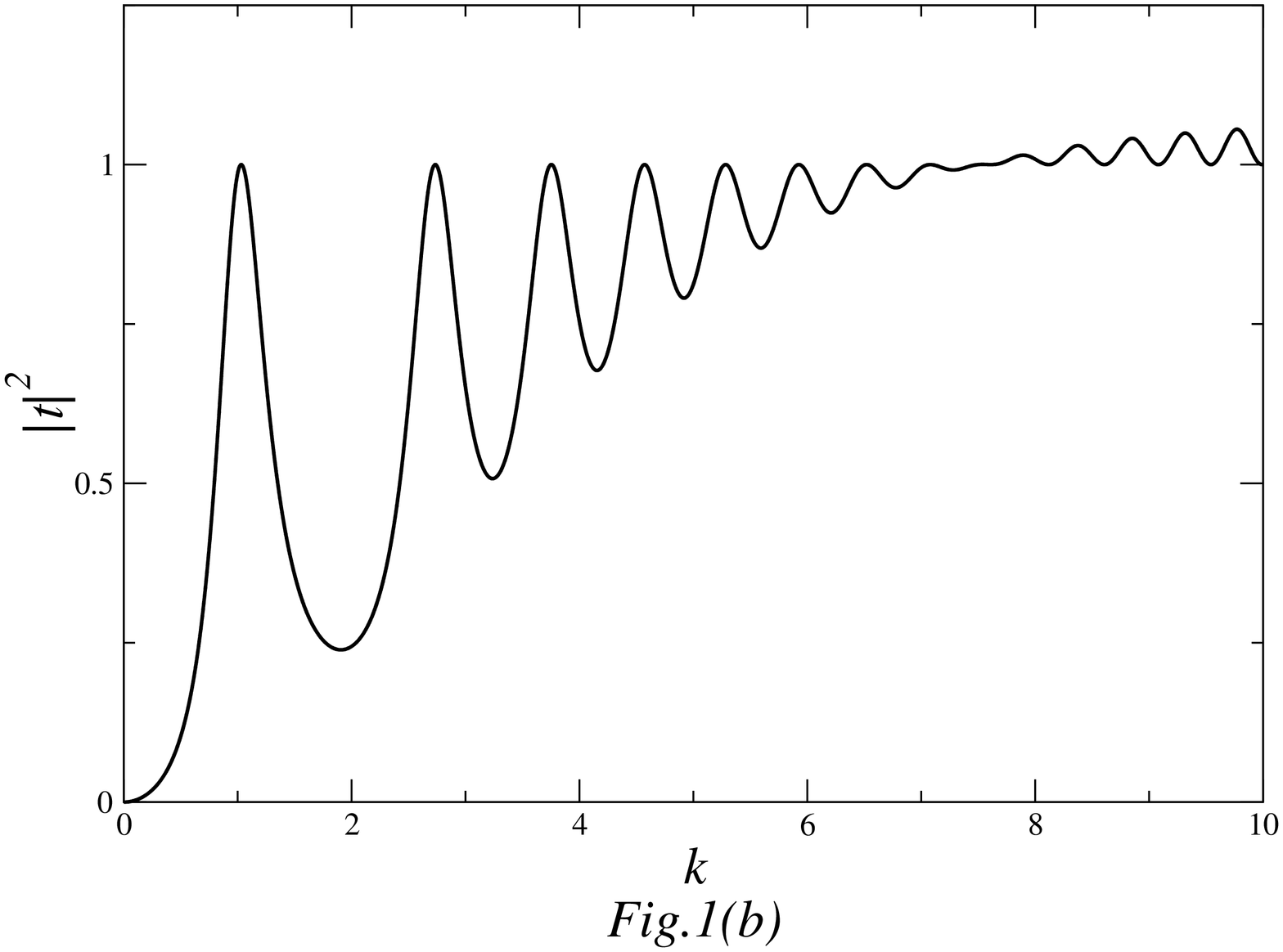}
\includegraphics*[width=2in,angle=270]{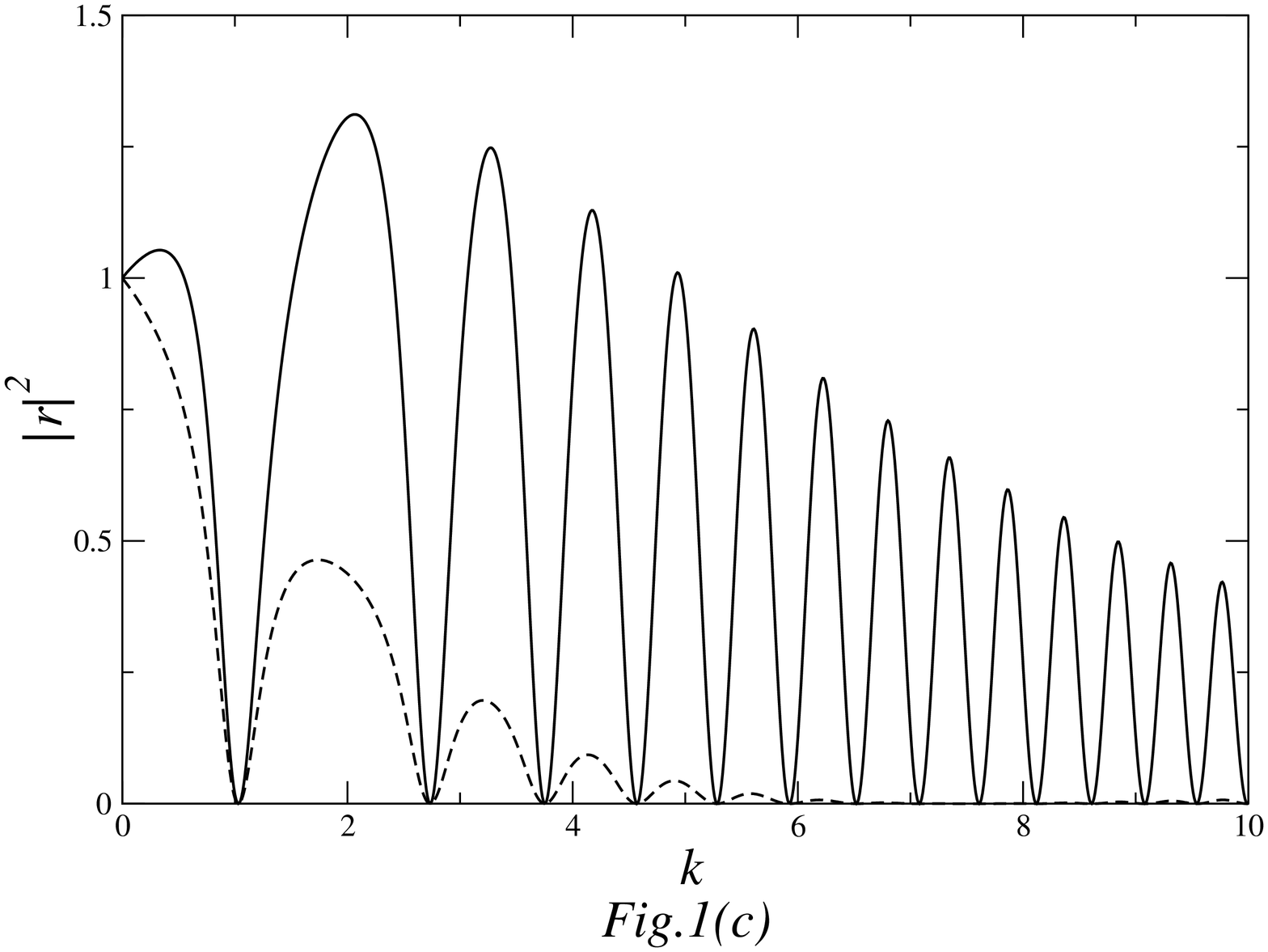}
\includegraphics*[width=2in,angle=270]{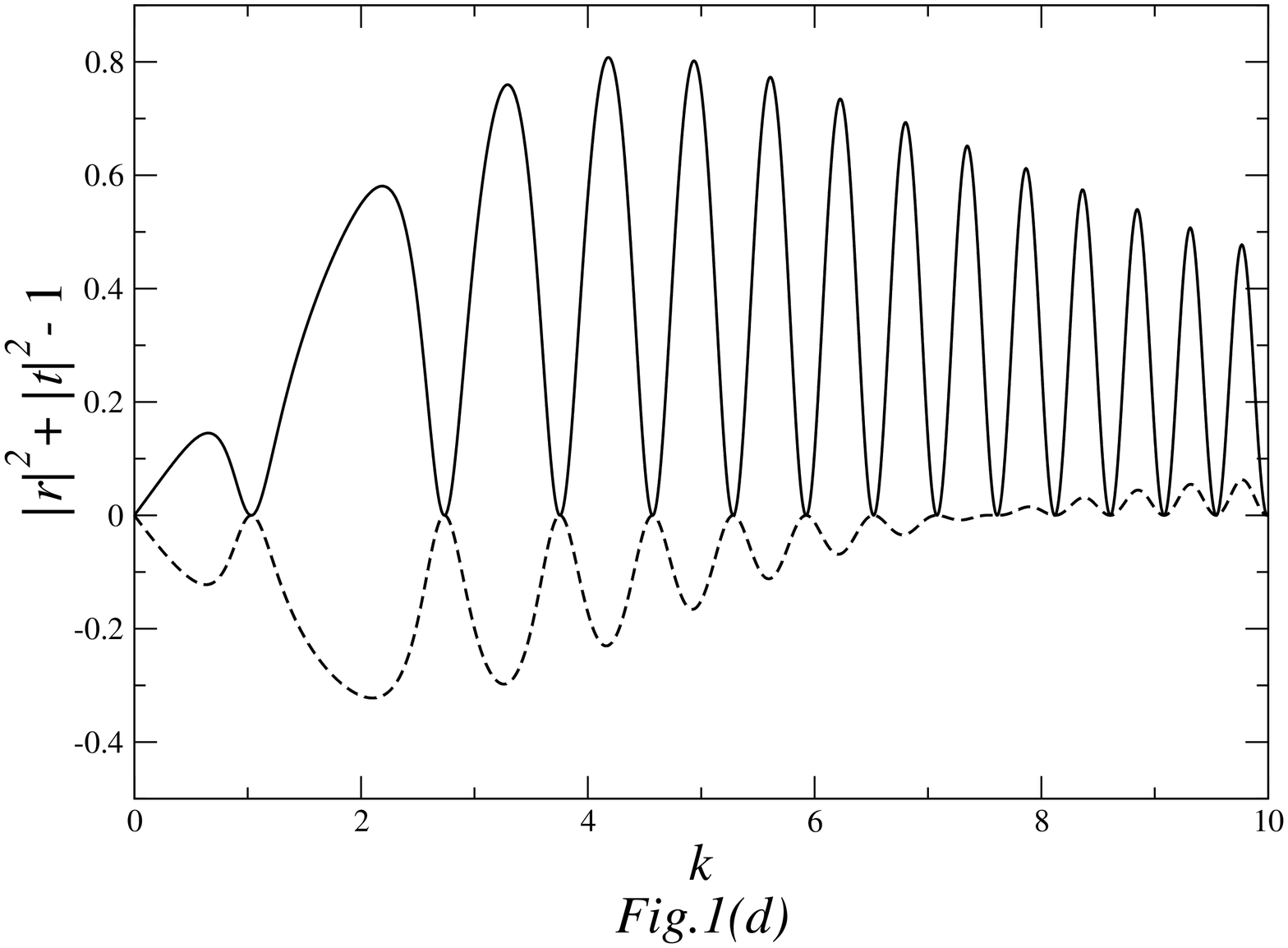}
\end{center}
\caption{The figures above are for Model-I. The solid lines are for left
incident beam and the dotted ones are for right incidence.
1(a): Variation of binding energy $(\beta =\sqrt{\frac{2mB}{\hbar^2}})$
with the strength of the imaginary part 
$(\tilde\lambda = \frac{2m\lambda}{\hbar^2})$ of the potential. Here we have 
taken $\frac{2mV_{0}}{\hbar^2} = 1$ and $a = 1$. 
1(b): Variation of transmission coefficient $(|t|^{2})$ with 
the wavenumber $(k)$ for 
left (or right) incident beam. 1(c): Variation of reflection coefficient 
 $(|r|^{2})$ with the wavenumber $(k)$ for left (or right) incident beam. 
1(d): Variation of the deviation
from unitarity $(|r|^{2} + |t|^{2} - 1)$ with the wavenumber $(k)$ for 
left (or right) incident beam.}
\label{fig.1 }
\end{figure}

\begin{figure}[t]
\begin{center}
\includegraphics*[width=2in,angle=270]{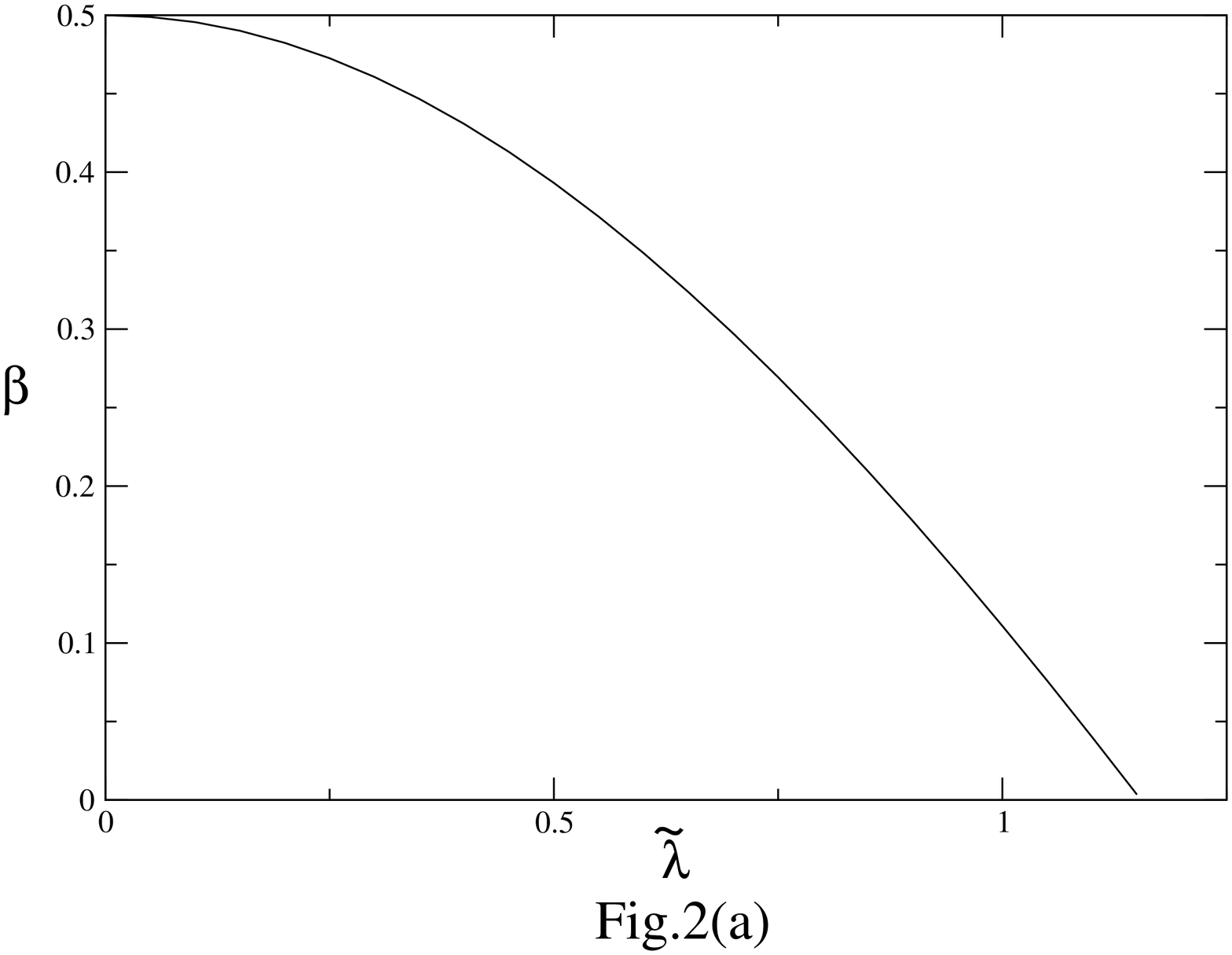}
\includegraphics*[width=2in,angle=270]{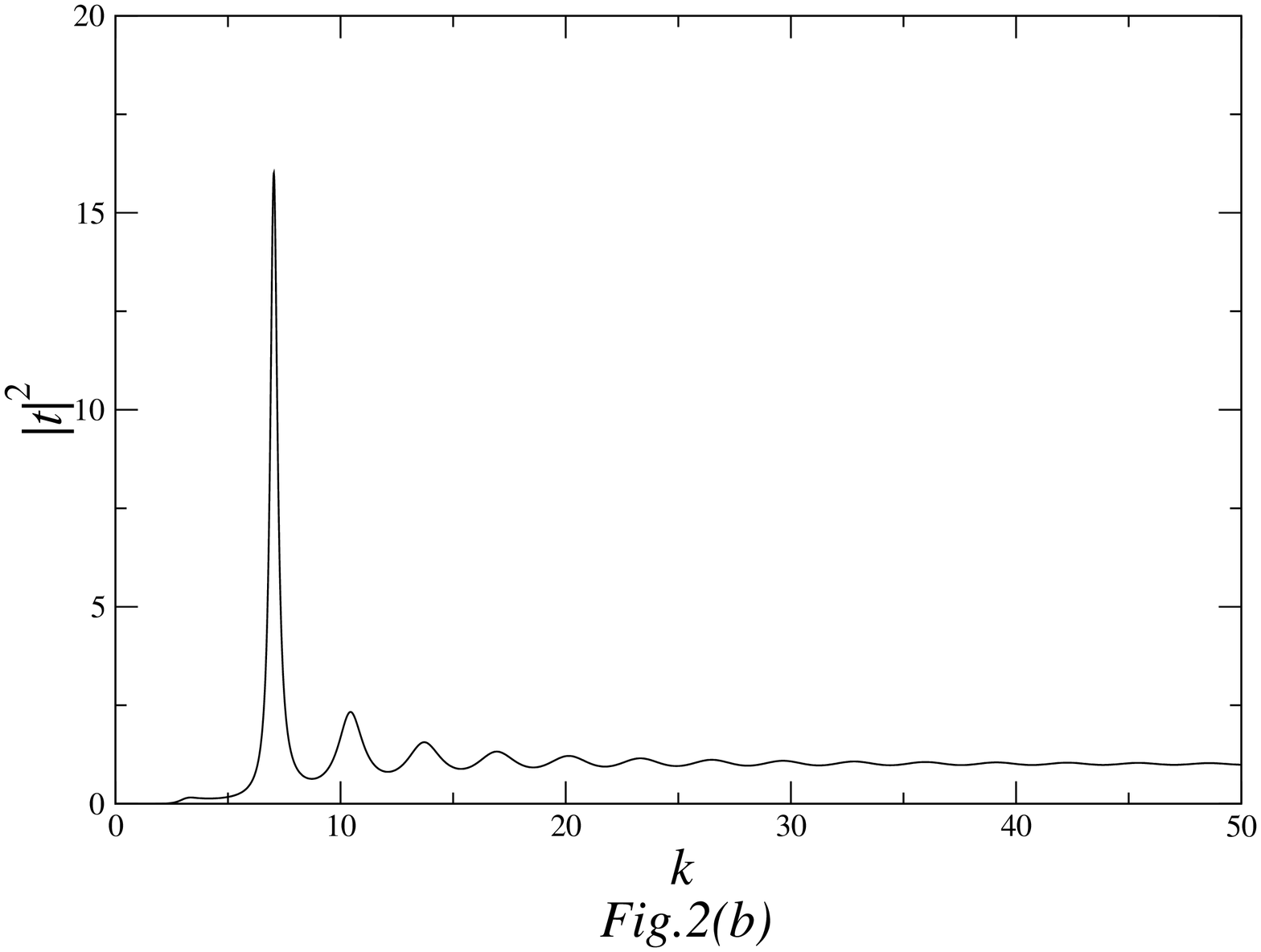}
\includegraphics*[width=2in,angle=270]{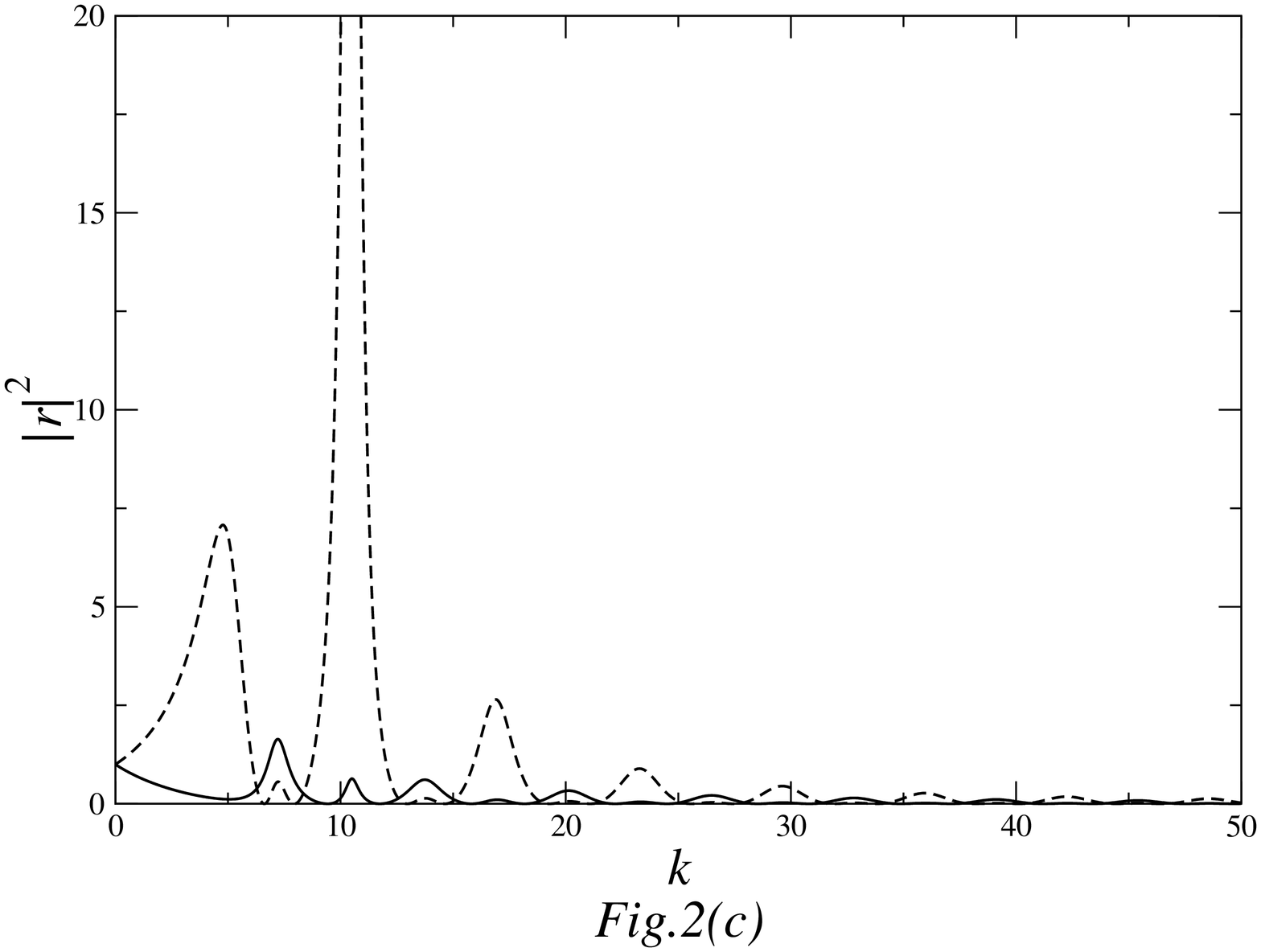}
\includegraphics*[width=2in,angle=270]{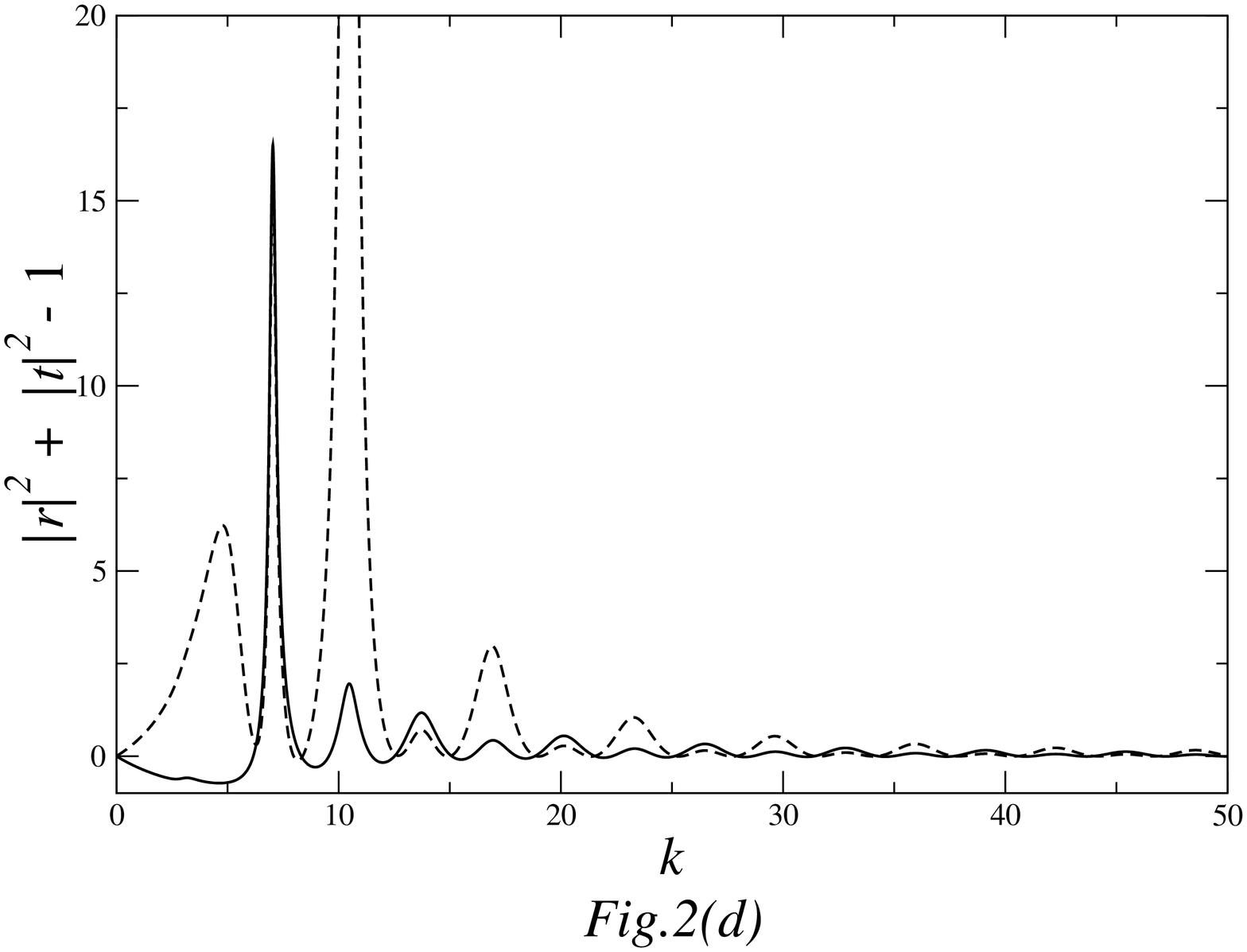}
\end{center}
\caption{The figures above are for Model-II. The solid lines are for left
incident beam and the dotted ones are for right incidence.
2(a): Variation of binding energy $(\beta =\sqrt{\frac{2mB}{\hbar^2}})$ 
with the strength of the imaginary part 
$(\tilde\lambda = \frac{2m\lambda}{\hbar^2})$ of the potential. 
Here we have taken $\frac{2mv_{0}}{\hbar^{2}} = 1$ and range $a = 1$. 
2(b): Variation of transmission coefficient $(|t|^{2})$ with the wavenumber 
$(k)$ for left
(or right) incident beam. 2(c): Variation of reflection coefficient $(|r|^{2})$
with the wavenumber $(k)$ for
left (or right) incident beam. 2(d): Variation of the deviation from unitarity
$(|r|^{2} + |t|^{2} - 1)$
with the wavenumber $(k)$ for left (or right) incident beam.}
\label{fig.2}
\end{figure}

For concreteness consider values of the potential parameters to be range
$a=10$, real part of potential given by $\frac{2mV_{0}}{\hbar^2}=100$ and the
strength of the imaginary part by $\frac{2m\lambda}{\hbar^2}=5$. We depict
the energy variation of the transmission coefficient (which is the same for
left and right incident beams) as a function of the wave number $k$ in 
$Fig.1(b)$. For the same potential the reflection coefficients for right and
left incident beams are depicted in $Fig.1(c)$. The quantity 
${{\vert}r{\vert}}^{2} + {{\vert}t{\vert}}^{2}$ which is equal to unity for 
real potentials will now depart from that value because of the imaginary part
and this is a measure of the extent of inelasticity. Accordingly we plot
${{\vert}r{\vert}}^{2} + {{\vert}t{\vert}}^{2}-1$ for both left and right 
incident beams in $Fig.1(d)$.

  As a second example of a $P$-Pseudo Hermitian Hamiltonian we choose a system
of a particle in one dimension governed by 
\begin{equation}
H=\frac{p^{2}}{2m} - v_{0}\delta(x) + i\lambda[\delta(x-\frac{a}{2})-\delta
(x+\frac{a}{2})].
\end{equation}
Here again $H^{\dagger}=PHP^{-1}$ but unlike case I the imaginary part of the 
potential is not proportional to the derivative of the real part. Defining
$\frac{2mv_{0}}{\hbar^2}=\tilde\mu$, $\frac{2m\lambda}{\hbar^2}=\tilde\lambda$
and $\frac{2mB}{\hbar^2}=\beta^2$ where $B$ is the binding energy, the 
eigenvalue condition obtained from the jump conditions for the wavefunction
becomes 
\begin{equation}
8\beta^3 - 4\tilde\mu\beta^2 = {\tilde\lambda}^{2}(1-e^{{-\beta}a})[\tilde\mu
(1-e^{-{\beta}a}) - 2\beta(1+e^{{\beta}a})].
\end{equation} 
Disregarding the unphysical double root of this equation at $\beta=0$ this is
solved numerically and the variation of the binding energy is shown as a 
function of the strength of the imaginary part of the potential in $Fig.(2a)$
where we have for illustration chosen the real part to be given by 
$\tilde\mu = \frac{2mv_{0}}{\hbar^2} = 1$ and the range $a=1$. It may
again be noted that there exists no bound state (in the usual meaning of the
word) below a certain critical 
value $(\mu_{cr})$ of the real part. This may easily be found from
the eigenvalue condition $[eq.(9)]$ by putting its non-trivial root equal to
zero and this leads to 
\begin{equation}
\mu_{cr} = \frac{4{\tilde\lambda}^{2}a}{4+{\tilde\lambda}^{2}a^{2}}.
\end{equation}
The transmission and reflection amplitudes for model-II are given by

\begin{equation}
t_L = t_R = {1 \over {(1-i\frac{\tilde\mu}{2k}) + (\frac{\tilde\lambda}{2k})^2
[-(1-i\frac{\tilde\mu}{2k}) - i\frac{\tilde\mu}{k}e^{ika} + 
(1+i\frac{\tilde\mu}{2k})e^{2ika}]}}, 
\end{equation}

\begin{equation}
r_L = {i\frac{\tilde\mu}{2k}(1-\frac{\tilde\lambda}{k} + 
\frac{{\tilde\lambda}^{2}}{2{k}^2}) + (1 - \frac{\tilde\lambda}{2k})
[(1 + i\frac{\tilde\mu}{2k})\frac{\tilde\lambda}{2k}e^{ika} - c.c.] 
\over {(1-i\frac{\tilde\mu}{2k}) + (\frac{\tilde\lambda}{2k})^2
[-(1-i\frac{\tilde\mu}{2k}) - i\frac{\tilde\mu}{k}e^{ika} +
(1+i\frac{\tilde\mu}{2k})e^{2ika}]}},
\end{equation}

\begin{equation}
r_R ={i\frac{\tilde\mu}{2k}(1+\frac{\tilde\lambda}{k}+
\frac{{\tilde\lambda}^{2}}{2{k}^{2}}) - (1 + \frac{\tilde\lambda}{2k})
[(1 + i\frac{\tilde\mu}{2k})\frac{\tilde\lambda}{2k}e^{ika} - c.c.] 
\over {(1-i\frac{\tilde\mu}{2k}) + (\frac{\tilde\lambda}{2k})^2
[-(1-i\frac{\tilde\mu}{2k}) - i\frac{\tilde\mu}{k}e^{ika} +
(1+i\frac{\tilde\mu}{2k})e^{2ika}]}},
\end{equation}
where $c.c.$ stands for the complex conjugate of the preceeding term in the 
braces.

The reflection and transmission coefficients for left and right incident beams
for this model and the departure from elastic unitarity are shown through
$Fig.(2b)$ to $(2d)$. For both the models considered it can be easily seen 
that the condition for poles of the reflection and transmission amplitudes is
exactly the equation for the binding energies, and thus the analytic structure
in the energy ($k^{2}$) plane of these amplitudes shows bound state poles.
They also have a cut along the real axis corresponding to a branch point at
$k^{2}=0$ but the discontinuity across the cut is provided by inelastic
unitarity arising from the imaginary part of the potential.  

Thus we have discussed pseudo-Hermitian Hamiltonians in the context of one
dimensional complex optical potentials and have shown that the energy 
eigenvalues are real for the strength of the imaginary part less than a 
critical value above which we obtain a  complex conjugate pair of eigenvalues 
which create some difficulties in their interpretation. We also discuss
transmission and reflection from such complex barriers and show that the 
reflection coefficient for left incident particles is different from that due
to those coming from the right. We believe it is important to gain experience
working with such pseudo-Hermitian Hamiltonians as their full physical
significance is not yet very clear.  

We gratefully acknowledge help from discussions with Swarnali Bandopadhyay and
Abhishek Choudhury.

\end{document}